\documentclass{elsart}

 \usepackage{graphicx}
  \include{epsfig}
\journal{Physica A}
\begin{document}
\begin{frontmatter}


\title{Assessing symmetry of financial returns series}
\author{H.F. Coronel-Brizio$^\dagger$},
\author{A.R. Hern\'andez-Montoya$^\dagger$\corauthref{cor}},
\ead{alhernandez@uv.mx}
\ead[url]{www.uv.mx/alhernandez}
\author{R. Huerta-Quintanilla$^+$},
\author{M. Rodr\'{\i}guez-Achach$^{\dagger+}$}

\corauth[cor]{Corresponding author: Maestr\'{\i}a en Inteligencia Artificial. Sebasti\'an Camacho 5, Xalapa Veracruz 91000, M\'exico. Tel/Fax: 52-228-8172957/8172855.}

\address{$^\dagger$ Facultad de F\'{\i}sica e Inteligencia Artificial.
Universidad Veracruzana, Apdo. Postal 475. Xalapa, Veracruz. M\'{e}xico}

\address{$^+$ Departamento de F\'{\i}sica Aplicada. Centro de Investigaci\'on y de Estudios Avanzados del IPN. Unidad M\'erida. Antigua carretera a Progreso km. 6, M\'erida, Yucat\'an, 97310, M\'exico}

\begin{abstract}
  \noindent
 Testing symmetry of a probability distribution is a common question arising from
applications in several fields. Particularly, in the study of observables
used in the analysis of stock market index variations, the question of symmetry
has not been fully investigated by means of statistical procedures.
In this work a distribution-free test statistic $T_{n}$ for
testing symmetry, derived by Einmahl and McKeague, based on the empirical likelihood approach, is used to address the study of symmetry of financial returns. The asymptotic points of the test statistic $T_{n}$ are also calculated and a procedure for assessing symmetry for the analysis of the returns of stock market indices is presented.
\end{abstract}

\end{frontmatter}




\vspace*{-.7cm}
 \section{Introduction}
\vspace*{-.9cm}
The gain/loss asymmetry of stock price variations  is considered as one of the stylized facts of financial time series \cite{cont} and its nature is of great and current interest \cite{Karpio}. In particular, and even though it has been researched for many years, the study of the symmetry of the unconditional distribution of financial returns remains as an important subject. For instance, in reference \cite{Lillo} conditions under which the distribution of ensemble returns becomes asymmetric are reported. On the other hand, \cite{peiro} has analized returns of a big sample of diverse financial indices without finding important symmetry deviations.\\

\vspace*{-.9cm}
Then, due to the importance of this subject, the assumption of symmetry of the distribution of
returns should be supported by means of objective distribution-free statistical procedures.\\

\vspace*{-.9cm}
In next section of this paper, we present and review a distribution-free test statistic $T_{n}$ for
testing symmetry, derived by Einmahl and McKeague \cite{Einhmal}, based on the empirical likelihood approach. In section \ref{ours} we show our numerical calculation of the asymptotic distribution of the $T_n$ statistic derived by simulation in \cite{Einhmal}. In section \ref{analisis} we present a procedure for assessing symmetry of returns distribution by using the statistic $T_n$ and illustrating it with data of the Mexican Stock Market Index IPC ({\it \'Indice de Precios y Cotizaciones} or by its English meaning Prices and Quotations Index) and the Dow Jones Industrial Average Index DJIA.\\

\vspace*{-1.1cm}
\section{The $T_{n}$ Statistic}
\vspace*{-.9cm}
\noindent
An approach to omnibus hypothesis testing based on the empirical likelihood method has been published
in a very interesting paper by Einmahl and McKeague \cite{Einhmal}. For testing the
null hypothesis of symmetry about zero, $H_{0}:F(0-x)=1-F(x-0)$, for all $x>0$
based on a sample $X_{1},\ldots,X_{n}$ of independent and identically distributed random variables
with common absolutely continuous distribution function $F$, they derived as a test statistic, the quantity:

\begin{center}
\begin{equation}
T_{n}=-2\int_{0}^{\infty} \log H(x)dG_{n}(x)=-\frac{2}{n}\sum_{i=1}^{n} \log H \left( \left| X_{i} \right| \right). 
\label{ts1}
\end{equation}
\end{center}

\noindent
$G_{n}$ denotes here the empirical distribution function of the $\left| X_{i} \right|$ and:
\vspace*{-.025cm}
%
\begin{eqnarray*}
 \log H(x)& = &nF_n \left( { - x} \right)\log \frac{{F_n \left( { - x} \right) + 1 - F_n (x - )}}
{{2F_n \left( { - x} \right)}} \\
  &+& n\left[ {1 - F_n \left( {x - } \right)} \right]\log \frac{{F_n ( - x) + 1
- F_n (x - )}}
{{2\left[ {1 - F_n \left( {x - } \right)} \right]}}, \\
 \end{eqnarray*}

\vspace*{-.05cm}
\noindent
where notation means $F_n(-x):=F_n(0-x)$ and $F_n(x-): = F_n(x-0)$.\\
\noindent 
The limiting distribution was found by proving that $T_n$ converges weakly to:
\begin{center}
\begin{equation}
 T_n \mathop  \to \limits^D \int\limits_0^1 { \frac{W(t)^2}{t} dt},
\end{equation}
\end{center}
\noindent
where $W$ denotes a standard Wiener process.\\

\section{Calculation of  the Asymptotic Distribution of $T_{n}$}
\label{ours}
\vspace*{-.9cm}
The asymptotic percentage points of the limiting distribution of $T_{n}$ were obtained here using (see for example \cite{Durbin}) the series representation:
\[
T_n \mathop  \to \limits^D \sum\limits_{i = 1}^\infty  {\lambda _i \nu _i }, 
\]
where $\nu_1,\nu_2,\ldots$ are independent chi-squared random variables, with one degree of freedom, and
$\lambda_1,\lambda_2,\ldots$ are the eigenvalues of the integral equation:

\begin{equation}
\int\limits_0^1 {\sigma (s,t)f_i ds = \lambda _i f_i (t)},
\label{inteq}
\end{equation}
\vspace*{-.15cm}
with $\sigma(s,t)$ denoting the covariance function of the process {\large $\frac{W(t)}{\sqrt{t}}$}.

\vspace*{-.1cm}
\noindent
Due to the difficulty of solving analytically equation (\ref{inteq}), the asymptotic percentage points of the distribution of $T_n$ were found numerically;  
using $k=100$ equally spaced points in the interval $(0,1)$
the integral was approximated in order to solve equation (\ref{inteq}). Similarly, a $k$ by $k$ grid on $(0,1)\times (0,1)$
was constructed to evaluate the covariance function $\sigma(s,t)$ and the eigenvalue problem solved to estimate $\lambda_1,\ldots,\lambda_k.$ Using these approximations, the asymptotic percentage
points were calculated using Imhof's method \cite{Imhof}. The above procedure was repeated for k=200 and k=300, and
the results compared. As it can be seen from table 1, the percentage points obtained are almost
identical except for a few discrepancies not greater than one unit in the third decimal figure. These
results are consistent with those obtained by simulation and reported in Einmahl and McKeague's paper.

\begin{table}[htb]
\begin{center}
\begin{tabular}{|c|c|c|}
\hline
Cumulative Probability &Percentage point ($k=300$)& Percentage point ($k=200$)\\ 
\hline
\hline
0.50&0.659&0.659\\
\hline
0.75&1.258&1.258\\
\hline
0.85&1.768&1.768\\
\hline
0.90&2.200&2.200\\
\hline
0.95&2.983&2.982\\
\hline
0.975&3.798&3.797\\
\hline
0.990&4.909&4.908\\
\hline
0.995&5.768&5.767\\
\hline
0.999&7.803&7.803\\
\hline
\hline
\end{tabular}
\caption{Asymptotic percentage points of $T_n$ calculated numerically. It can be seem from two columns values that numerical convergence of $T_n$ is very fast.}
\label{cp}
\end{center}
\end{table}
\vspace*{-.7cm}
\section{Proposed approach and examples}
\label{analisis}
\vspace*{-.9cm}
Given a set of observations from an unknown probability distribution, if the symmetry
point is known, a statistical procedure (as the one described above) can be used to
test the symmetry of the distribution around that point. However, when the symmetry point is unknown, it
might happend that the test would lead us to the rejection of this assumption,
even when the distribution is symmetric; this would be the case when the symmetry
point is incorrectly specified in the test.\\
\noindent
Let us denote by $\{S_{t}\}$ the stock index process and by 
$R_{t}=\log{S_{t}}-\log{S_{t-\Delta t}}$ its returns or logarithmic increments during a certain time interval $\Delta t$. The ``shifted returns'' are also defined as
$R_{t}(c)=R_{t}-c$, where $c$ denotes a real number. Finally, let us
denote by $T_{n}(c)$ the value of the test-statistic $T_{n}$ calculated
from $R_{1}(c),\ldots,R_{N}(c)$ for a particular value of $c$.\\
\noindent
In the following, we will mean by a {\em plausible} value of the symmetry
point, (for a significance level $\alpha$) any real number $c_{0}$, such that
$T_{n}(c_{0})<T(\alpha)$ where $T(\alpha)$ denotes the $\alpha-$level
upper point of the distribution of $T_{n}.$

Using a similar approach to that of constructing confidence regions, a plot of
$T_{n}(c)$ versus $c$ can be used to identify a {\em plausible} set of values of
the unknown symmetry point $c$ in the sense that, for a given significance level $\alpha$,
the interval would contain the set of all possible values of $c$ which would not lead
to the rejection of the null hypothesis of symmetry for the probability distribution of the
random variable $R_{t}$.\\
\noindent
In order to illustrate the procedure, we present our analysis for two data sets:
\begin{enumerate}
\item DJIA Daily closing values from October 30, 1978 to October 20, 2006.
\item IPC Daily closing values  for the same period.
\end{enumerate}
\noindent
For each data set, the shifted returns $R_{t}$ were obtained, and the plots produced using the procedure described above.\\
\noindent
In figure 1, it is shown the plot from the Dow Jones index data, including
the lines $y=4.909$, $y=2.983$ and $y=2.200$, which correspond to the asymptotic 0.99, 0.95
and 0.90 percentiles of distribution of the $T_{n}$ statistic, from table 1. As it can be seen, for
a significance level $\alpha=0.10$ (or lower), it is possible to find an interval of
plausible values for the unknown point of symmetry which would not lead us to the rejection of
the assumption of symmetry. Approximately, for $\alpha=0.10$, any value within the interval
$(2.6\times 10^{-4},6.2\times 10^{-4})$ can be statistically considered as a point around which
the distribution of the returns is symmetric.

\begin{figure}[htb!]
\begin{center}
\resizebox{0.55\textwidth}{!}{%
\includegraphics{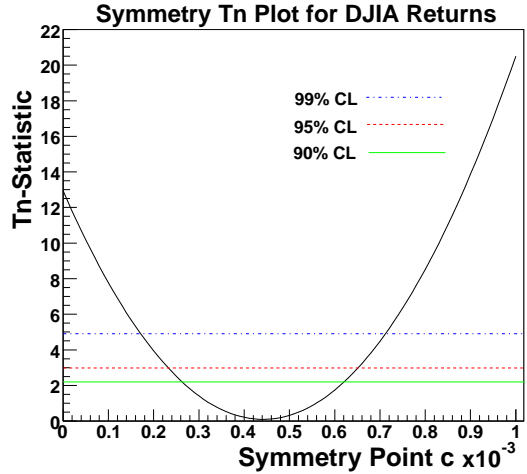}}
\caption{Plot of statistic $T_{n}(c)$ versus selected values
of the symmetry point $c$ for the Dow Jones return series data.
Horizontal straight lines correspond to the 99, 95 and 90 upper percentage points, as indicated}
\label{djplot1}
\end{center}
\end{figure}
\begin{figure}[htb!]
\begin{center}
\resizebox{0.55\textwidth}{!}{%
\includegraphics{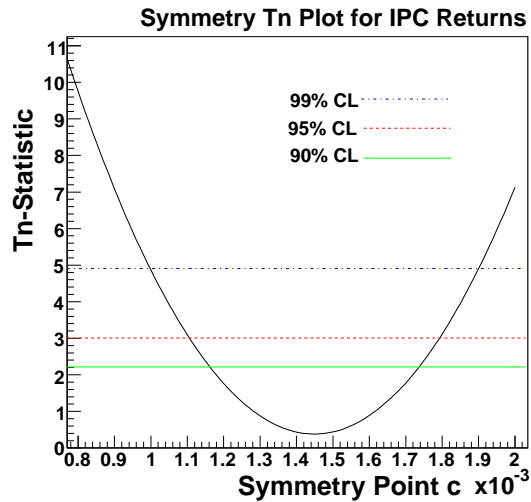}}
\caption{Plot of statistic $T_{n}(c)$ versus selected values
of the symmetry point $c$ for the IPC return series data.
Horizontal straight lines correspond to the 99, 95 and 90 upper percentage points.}
\label{ipcplot1}
\end{center}
\end{figure}

\noindent
Figure 2 shows the symmetry plot for the returns obtained from the Mexican IPC index data.
Considering the 90\% percentage line, we find that an
interval of plausible values for the point of symmetry can be found; approximately
the interval $(1.16\times 10^{-3}, 1.74\times 10^{-3})$ would be a $90\%$ confidence-interval for
the unknown point of symmetry; that is, if we choose any value for the symmetry point within that
interval, the statistic $T_{n}$ would not lead to the rejection of the hypothesis of symmetry
around the choosen point. Again, our assessment would be that, for a given significance level $\alpha=0.10$ (or lower), there exists a set of plausible values for which the assumption of symmetry can be statistically
supported.

It must be remarked that the approach discussed here is not equivalent to that of maximizing
a test-statistic as it has been the case, for example, in \cite{Karsten} or \cite{Coronel_tesis} and \cite{Coronel}. The reasoning behind our assessment is based on the idea that
whenever there exists a plausible value for the point of symmetry, this assumption can be
statistically sustained. 

\vspace*{-.7cm}
\section{Conclusions}
\vspace*{-.9cm}

A procedure for assessing the assumption of symmetry, for the probability distribution
function of returns, has been presented. The approach is based on determining, statistically, whether
or not, a set of plausible values for the unknown symmetry point can be found. Two examples were
discussed to illustrate the approach, analyzing returns data from the Dow Jones and the Mexican IPC stock market indices. In both cases, sets of plausible values for the point of symmetry could be
found, so that that the assumption of symmetry can be statistically supported.

{\bf Acknowledgments\\}
\noindent
The authors wish to thank professors Einmahl and McKeague, for kindly allowing
the use of their computer routines for calculating $T_{n}$. We appreciate the valuable suggestions from N. Cruz, P. Giubellino, S. Jim\'enez, E. Rojas and R. Vilalta. We also are very grateful to P. Zorrilla-Velasco, A.Reynoso-del Valle and S. Herrera-Montiel, from the BMV for providing us with the IPC data and their valuable time and cooperation. \\
\noindent
This work has been supported by Conacyt-Mexico under Grants 44598
and 45782. Plots have been done using ROOT \cite{root}.


\vspace*{-.7cm}
\begin{thebibliography}{}
\vspace*{-.7cm}

\bibitem{cont} R. Cont. Quantitative Finance 1 (2001) 223-236. 
\vspace*{-.2cm}
\bibitem {Karpio} K. Karpio, M. A. Zaluska-Kotur, A. Orlowski, Physica A 375 (2007) 599-604.
\vspace*{-.2cm}
 \bibitem{Lillo} F. Lillo and R.N. Mantegna, Eur. Phys. J. B 15 (2000) 603-606. 
\vspace*{-.2cm}
\bibitem{peiro} A. Peiro. Quantitative Finance 4 (2004) 37-44. 
\vspace*{-.2cm}
\bibitem{Einhmal} H.J. Einmahl and I.W. McKeague, Empirical likelihood based hypothesis testing. Bernoulli 9 (2003) 267-290. 
\vspace*{-.2cm}
\bibitem{Durbin} Durbin, J. {\em Regional Conference Series in Appl. Math.},{\bf 9}, (1973) Philadelphia: SIAM.
\vspace*{-.2cm}
\bibitem{Imhof} Imhof, J.P. Biometrika, 48 (1961) 419-426. 
\vspace*{-.2cm}
 \bibitem{Karsten} Karsten, P. The generalized hyperbolic model: estimation,
financial derivatives and risk measures, Ph.D. Thesis, Freiburg University 1999.
\vspace*{-.2cm}
 \bibitem{Coronel_tesis} H.F. Coronel-Brizio, Regression tests of fit and some comparisons. Ph.D. Thesis. Department of Mathematics and  Statistics, Simon Fraser University 1994. 
\vspace*{-.2cm}
 \bibitem{Coronel} H.F. Coronel-Brizio, A.R. Hernandez-Montoya, Physica A
  354 (2005) 437-449.
\vspace*{-.2cm}
\bibitem{root}  Nucl. Inst. \& Meth. in Phys. Res. A 389 (1997). http://root.cern.ch.


\end{thebibliography}
\end{document}